\documentclass[amsmath]{JHEP3}

\usepackage{amsmath}
\usepackage{amssymb}
\usepackage[T1]{fontenc}
\usepackage{longtable}
\usepackage{multicol}
\usepackage{cite}
\usepackage{epsfig}

\newcommand{\CO}{{\ensuremath \cal O}}
\newcommand{\CR}{{\ensuremath \cal R}}
\newcommand{\msub}[1]{\ensuremath _{\mbox{\scriptsize #1}}}
\newcommand{\be}{\begin{equation}}
\newcommand{\ee}{\end{equation}}

\def\lsi{\raise0.3ex\hbox{$<$\kern-0.75em\raise-1.1ex\hbox{$\sim$}}}
\def\gsi{\raise0.3ex\hbox{$>$\kern-0.75em\raise-1.1ex\hbox{$\sim$}}}

\title{A comprehensive search for the $\Theta^+$ pentaquark on the  lattice}

\author{F.~Csikor$^a$, Z.~Fodor$^{a,b}$, 
        S.D.~Katz$^a$,   T.G.~Kov\'acs$^c$ 
   and B.C.~T\'oth$^a$  \\ 
    $^a$Institute for Theoretical Physics, E\"otv\"os University, Hungary \\
    $^b$Department of Physics, University of Wuppertal, Germany\\
    $^c$Department of Theoretical Physics, University of P\'ecs, Hungary }

\abstract{
We study spin 1/2 isoscalar and isovector, even and odd parity candidates
for the $\Theta^+(1540)$ pentaquark particle using large scale lattice QCD 
simulations. Previous lattice works led to inconclusive results because
so far it has not been possible to unambiguously identify the known  
scattering spectrum and tell whether additionally a genuine pentaquark 
state also exists. Here we carry out this analysis using several
possible wave functions (operators).  Linear combinations of those 
have a good chance of 
spanning both the scattering and pentaquark states. 
Our operator basis is the largest in the 
literature, and it also includes spatially non-trivial
ones with unit orbital angular momentum.
The cross correlator we compute is 14$\times$14 with 60 
non-vanishing elements.
We can clearly distinguish the lowest scattering state(s)
in both parity channels up to above the expected location of
the pentaquark, but we find no trace of the latter.
Based on that we conclude that there are most probably 
no pentaquark bound states at our quark masses, corresponding to 
$m_\pi$=400--630~MeV. However, we cannot rule out the existence
of a pentaquark state at the physical quark mass corresponding to
$m_\pi$=135~MeV or pentaquarks with a more exotic wave function. 
}

\preprint{ITP-Budapest-619 \\ WUB-05-02}

\begin{document}

\section{Introduction}
     \label{se:Intro}

One of the mysteries of hadronic physics has been the failure to observe 
baryon states with quantum numbers that cannot be explained in terms 
of three quarks. However, for a long time  this was  not considered to be 
a practical problem due to the  presumed large decay width of these exotic 
baryons. The experimental signal of the 
$\Theta^+(1540)$ particle \cite{Nakano:bh},
\cite{Stepanyan:2003qr}-\cite{Troyan}  changed  
this situation dramatically.
Indeed, the  experimental upper bound so far 
on the width of the $\Theta^+$
is around 10~MeV. This remarkably narrow width 
would also explain why the $\Theta^+$ has not been seen before.
Since the $\Theta^+$ was observed to decay into a neutron and a 
$K^+$, its strangeness has to 
be  +1, the third component of its isospin is 0, and its minimal quark content 
is $dduu\bar{s}$. From the lack of a signal in the $I_3\!=\!1$ channel
the SAPHIR collaboration concluded that the $\Theta^+$ is most
probably an isospin
singlet state \cite{Barth:2003es}. Its spin and parity cannot be 
pinned down based on currently available experimental data. 

Though the $\Theta^+(1540)$ is seen experimentally in low energy exclusive 
processes, there are a number of ($e^+e^-$ or 
high-energy proton collision) experiments, where the $\Theta^+(1540)$ 
is not seen \cite{Bai:2004gk} --\cite{Armstrong:2004gp}. The different 
kinematical and experimental conditions between the low energy 
exclusive experiments (with experimental evidence for $\Theta^+$)
and the inclusive experiments (usually non-observations) do not 
allow a direct comparison so that the null results do not prove that the 
positive experiments are wrong \cite{Rossi:2004rb}. Nevertheless, it is 
fair to say that the experimental situation is not perfectly 
clear at the moment. Since there are only single experimental indications
of other exotic pentaquarks (the possible
$\Xi^{--}(1860)$ state reported by the NA49 experiment at
CERN~\cite{Alt:2003vb} and the charmed pentaquark identified
by the H1 experiment at DESY  \cite{Aktas:2004qf})
their existence is even more debated than that of the $\Theta^+(1540)$.

Originally, the experimental search for the $\Theta^+(1540)$ was largely 
motivated by the chiral soliton model \cite{soliton} 
that predicted for the first time in 1997 a  
mass of 1530~MeV and a width of less than 15 MeV for this exotic S=+1 
baryon  (for an earlier estimate of the mass in the
soliton approach see \cite{Praszalowicz:2003ik}). 
The experimental evidence of the $\Theta^+$ pentaquark triggered a flurry of
theoretical speculations about its possible structure, yet unmeasured
quantum numbers and on the possibility of the existence of other
exotic hadrons. A particularly popular and successful approach is based on 
different types of quark models 
\cite{Jaffe:2003sg,diquark,quarkmodel}. 
Attempts have been made to understand the experimental
findings by means of baryon-meson bound states \cite{bound} as well as
QCD sum rules \cite{sumrules}.

These models substantially differ in the properties they predict for the
pentaquark state. E.g.\ several models predict 
positive parity, while other approaches insist on negative parity. 
Clearly, it is of utmost importance to study the 
$\Theta^+(1540)$ without any model assumptions, based on a first 
principles non-perturbative approach, i.e.\ lattice QCD.

The difficulty of the lattice approach 
lies in the fact that the $\Theta^+(1540)$ mass is very close 
to the NK scattering threshold.  In lattice QCD one has to use a finite 
box, implying that the continuum of KN scattering states turns into
a stack of discrete energy levels with the $\Theta^+(1540)$ embedded 
somewhere among them. It is then not an easy task to reliably distinguish the 
$\Theta^+(1540)$ from these nearby scattering states 
since all the quantum numbers coincide. 

There are a few published works on the $\Theta^+(1540)$ in lattice QCD. 
Considering the difficulties involved, it is not surprising
that the results are not in complete agreement.
Here we collected the main features of these
studies, for a more detailed discussion see \cite{review}.
Except for one, all the lattice studies report a signal in the negative 
parity channel close to the expected location of the pentaquark.  
Based on the simple fact that the lowest state with opposite
parity lies much higher, Refs.\ \cite{Csikor:2003ng,Sasaki:2003gi}
tentatively identify this state with the $\Theta^+$. Others
employ finite volume analysis \cite{Mathur:2004jr} and 
twisted boundary conditions \cite{Takahashi:2004sc} to distinguish
between a two-particle and a one-particle state and they conclude
that what they see is a scattering state. Ref.\ \cite{Takahashi:2004sc}
on the other hand identifies the first excited state with negative 
parity and from its dependence on the volume concludes that it is
the $\Theta^+$ resonance. All these works are
largely consistent in the lowest masses they find in both parity
channels, they only differ in their interpretations. 

The only result, which is inconsistent with the rest is that of 
\cite{Chiu:2004gg}, observing a state in the positive parity channel
compatible with the $\Theta^+$ and a much higher state in the 
negative parity channel. We stress that none of the lattice studies
so far could identify the lowest expected scattering state in {\em both}
parity channels. This strongly suggests that the wave functions,
all based on rotationally symmetric quark sources at the origin,
do not have sufficient overlap with all the low lying states. 
Another lattice study, Ref.\ \cite{Alexandrou:2004},
finds some evidence that a pentaquark potential based on the
diquark-diquark-antiquark picture is energetically more 
favourable than that of the KN picture.
All these investigations use the quenched approximation, but with 
different fermion formulations and pentaquark wave functions (operators). 

A reliable confirmation of the existence of the 
$\Theta^+(1540)$ from lattice studies is achieved only if 
all the states up to above the expected location of the $\Theta^+(1540)$
have been identified and the $\Theta^+(1540)$ can be clearly distinguished
from the neighbouring scattering states.
It is thus clear that a further more comprehensive study is required
and this is our aim in the present paper. 

Here we use several possible wave functions (operators) 
that have a good chance of spanning both the scattering and 
pentaquark states. Our operator basis is the largest in the 
literature, the cross correlator we compute is 14$\times$14 with 60 
non-vanishing elements. In particular we also include displaced, rotationally
non-symmetric spatial quark configurations to allow non-zero 
orbital angular momentum as well as a better separation
of the scattering states.

In the positive parity channel the lowest 
state we can identify is compatible with the lowest expected 
two-particle state and is already significantly above the $\Theta^+$.
In the negative parity channel we can distinguish the two lowest
states that both turn out to be compatible with the expected scattering 
states. At the box volumes we use the $\Theta^+$ is expected to
be between the two lowest scattering states, but we see no trace
of it there. We also carried out the analysis for a smaller 
volume and found that the volume dependence of the energies
is compatible with all the identified states being two-particle states.

In conclusion, we identified all the states around the expected 
location of the $\Theta^+$ in both parity channels and they all
turned out to be significantly different from the $\Theta^+$.
Since our $u$ and $d$ quarks were heavier than the physical
quarks (corresponding to $m_\pi$=400--630~MeV) we cannot rule
out the possible appearance of a pentaquark state for lighter
quarks. Although not very likely, it is also possible that 
a pentaquark state exists with a wave function having very
small overlap with all our trial wave functions.


In the present study we chose to work
in the quenched approximation again, which is known to be quite successful
in reproducing mass ratios of stable hadrons
\cite{Gattringer:2003qx} --\cite{Shanahan:1996pk}. Compared to our
previous analysis we improved by three means. In addition to the
cross correlator technique and the finite volume analysis we
increased our statistics by a factor of 2--3.

\section{Cross correlators}

In hadron spectroscopy one would like to identify states with 
given quantum numbers by computing the vacuum expectation 
value of the Euclidean correlation function 
$\langle 0| \CO(t) \bar{\CO}(0) |0 \rangle$ of some composite hadronic
operator $\CO$. The operator $\CO$ is built 
out of quark creation and annihilation operators.
In physical terms the correlator is the amplitude 
of the ``process'' of creating a complicated hadronic state 
described by $\CO$ at time $0$ and destroying it at time $t$.

After inserting a complete set of eigenstates 
$|i \rangle$ of the full QCD Hamiltonian the correlation function
can be written as
\begin{equation}
  \langle 0| \CO(t) \bar{\CO}(0) |0 \rangle = 
  \sum_i \;\; |\; \langle i | \bar{\CO}(0) |0 \rangle \; |^2 \; 
   \; \mbox{e}^{-(E_i-E_0)t},
     \label{eq:corr}
\end{equation}
where 
\begin{equation}
  \CO(t) = \mbox{e}^{-Ht}\; \CO(0) \; \mbox{e}^{Ht}
\end{equation}
and $E_i$ are the energy eigenvalues of the Hamiltonian. 

Note that since we work in Euclidean space-time 
(the real time coordinate $t$ is replaced with $-it$), the
correlators do not oscillate, they rather die out exponentially
in imaginary time. In particular, after long enough time 
only the lowest state created by $\CO$ gives 
contribution to the correlator. The energy eigenvalue 
corresponding to that state can be extracted from an exponential
fit to the large $t$ behaviour of the correlator.

In principle higher states could also be identified 
by generalizing the procedure and fitting the correlator with
a sum of exponentials. In practice, however, that would require
extremely high precision data, usually not available in lattice
simulations. A much more realistic solution can be based on the
observation that if the operator $\CO$ happened to have
negligible overlap with the ground state in the given
sector, a single exponential fit would yield the first excited
state. This, however, is very unlikely to happen by sheer luck,
as it would require fine tuning. 

It is exactly this fine tuning that can be performed if instead
of one operator $\CO$ one considers a linear combination of the form 
\begin{equation}
 \CR(t) = \sum_{i=1}^n v_i \CO_i(t).
\end{equation}
The correlator of $\CR$ can be easily expressed in terms of the $n\times n$
correlation matrix 
\begin{equation}
 C_{ij}(t) = \langle \CO_i(t) \bar{\CO}_j(0) \rangle 
\end{equation}
as
\begin{equation}
 R(t) \; = \; \langle \CR(t) \bar{\CR}(0) \rangle \; = \; 
    \sum_{i,j=1}^n v_i \bar{v}_j C_{ij}(t).
\end{equation} 
Morningstar and  Peardon used this cross correlator
to compute glueball masses on the lattice \cite{Morningstar:1997ff}.
Their procedure was based on the effective mass defined for a general
correlator as
\begin{equation}
 m\msub{eff} \; = \; -\frac{1}{\Delta t} 
                      \ln\left( \frac{C(t+\Delta t)}{C(t)}\right).
\end{equation} 
Let us now consider the effective mass obtained from $R(t)$,
\begin{equation}
 m(t) = -\frac{1}{\Delta t}\ln\left[\frac{R(t+\Delta t)}{R(t)}\right] = 
        -\frac{1}{\Delta t} 
            \ln\left[ \frac{\sum_{i,j=0}^n v_i \bar{v}_j C_{ij}(t+\Delta t)}{
                        \sum_{i,j=0}^n v_i \bar{v}_j C_{ij}(t)}\right].
   \label{eq:effmass} 
\end{equation}
This can be exploited to construct linear combinations
that have optimal overlap with the ground state or higher excited states.
If the correlator contained only $n$ different states, the linear
combination with the lowest effective mass would yield exactly the
ground state. In practice this is a good approximation starting already
from moderate values of $t$, since higher states die out rapidly.

A simple
computation shows that the stationary points of the effective mass
with respect to the variables $\{v_i\}_{i=1}^n$ are given by the solutions
of the generalized eigenvalue equation
\begin{equation}
 \sum_{i=1}^n C_{ij}(t+\Delta t) v_j \; = \;
 \lambda \sum_{i=1}^n C_{ij}(t) v_j.
\end{equation}
    %
    %
    %

Initially we only asked for the lowest effective mass, but this
eigenvalue problem can have many solutions. It is not hard to interpret
them using the following geometric picture. $C_{ij}(t)$ and 
$C_{ij}(t+\Delta t)$, being both Hermitian, can be considered to
be the components of two quadratic forms on the n-dimensional 
space spanned by the $v_i$'s. Let us interpret $C_{ij}(t)$ as 
an inner product on this vector space. 
It can be seen from eq.\ (\ref{eq:effmass})
that the effective mass does not depend on the normalization
of the vector $\{v_i\}$, so we can restrict it to be of unit length
(with respect to the inner product just defined). It is now easy 
to see that the stationary points of the effective mass correspond
to the principal axes of the second quadratic form, $C_{ij}(t+\Delta t)$.
In the language of the generalized eigenvalue problem this is 
equivalent to the statement that two quadratic forms can always be 
simultaneously diagonalized in a vector space: there is a basis
orthonormal with respect to one quadratic form and pointing along
the principal axes of the other one. 

Assuming a generic case with no degeneracies the stationary points
will have 0,1,2... unstable directions and they yield the coefficients
of the linear combinations 
corresponding to the ground state and the higher excited states. Of course
this statement again is exactly true only if there are only
$n$ states in the correlator. The importance of corrections
coming from higher states can be estimated by checking
how stable the whole procedure is with respect to varying 
$t$ and $\Delta t$. 

This gives a general method to determine the
optimal linear combinations of $n$ operators that have the best
overlap with the lowest $k$ $(k\leq n)$ states. The only disadvantage
of this procedure is that being based on effective masses, it 
always uses only two points of the correlators to extract the
optimal linear combinations. On the other hand, once the optimal 
linear combinations have been found the corresponding correlators
can be fitted using any standard technique.

\section{Details of the simulation}
\subsection{Choice of operators}
One of the most important parts of the whole analysis is the proper choice 
of operators. We need a large number of independent operators, which span a 
large enough subspace containing the scattering states and a possible 
pentaquark state. 

In order to have really independent operators, 
we used non-trivial wavefunctions
for the quark fields. The typical operators used in hadron spectroscopy 
contain quarks at only one lattice point with some Gaussian smearing.
These operators have automatically zero orbital angular momentum and a
spin eigenstate can be guaranteed by correctly choosing the Dirac-structure 
of the operator. This, however, gives a very limited set of operators.
Moreover, some operators, e.g.\ the one proposed by Jaffe and 
Wilczek \cite{Jaffe:2003sg} cannot be implemented in this way.

Therefore we decided to use operators, which contain quark fields at
different lattice sites. In general the five-particle wave function could
be any function of the locations of the five quarks. However, since the 
correlation functions are built up from quark propagators, we have to restrict
ourselves to wave functions, which are products of the individual quark wave 
functions:
\be
O(x_1,x_2,x_3,x_4,x_5)=q_1(x_1)q_2(x_2)q_3(x_3)q_4(x_4)q_5(x_5)
\ee

Here, for simplicity we omitted the color and Dirac-structure.
These are the elementary operators for which the correlators can be computed 
by single Dirac-matrix inversions. A general five-quark operator can be 
written as a linear combination of such elementary operators.

For the individual quark wave-functions we use a simple Gaussian function
centered at some lattice site:
\be
q_i(x_i)=\exp\left(-\frac{(x_i-x_{i0})^2}{r_i^2}\right).
\ee

It is easy to see that if not all $x_{i0}$-s are the same then the operator 
will not have a spherical symmetry and therefore it will create a mixture 
of angular-momentum eigenstates. 
According to Appendix \ref{ap:spin_proj} we can 
project out angular momentum 1/2 using the projector $P^{(G_1)}$.

We had two sets of operators, one with spatially completely symmetric and
one with antisymmetric operators. Since the cross-correlator of a symmetric
and antisymmetric operator vanishes we could perform the runs separately for
the two sets. It turned out that 
the symmetric/antisymmetric operators had a good overlap 
with negative/positive parity states, respectively.

Let us simplify our notation further by allowing only quark wave-functions 
that are centered on points of the $z$ axis only. Operators based on such 
wave-functions have axial symmetry and therefore the spin-projection requires
a minimal number of extra operators. Let
\be
q_i(d_i,r_i,x_i)=\exp\left(-\frac{(x_i-d_i\cdot\hat{z})^2}{r_i^2}\right),
\ee
where $\hat{z}$ is the unit vector along the $z$ axis. We will usually omit the
$x_i$ argument.

We used the following set of isoscalar operators:
\begin{eqnarray}
\CO_1&=&\epsilon^{abc}[u_a^T(0,4) C\gamma_5 d_b(0,4)] 
           \{u_c(0,4) \bar{s}_e(0,4)\gamma_5 d_e(0,4) + (u \leftrightarrow d) \} \nonumber \\
\CO_2&=&\epsilon^{abc}\epsilon^{ade}\epsilon^{bgh}[u_d^T(0,4) C\gamma_5 d_e(0,4)]
	[u_g^T(0,4) C d_h(0,4)] C\bar{s}_c^T(0,4)  \nonumber \\
\CO_3&=&P^{(G_1)} \left[ \epsilon^{abc}[u_a^T(0,4) C\gamma_5 d_b(0,4)] 
           \{ u_c(0,4) \bar{s}_e(N_s/2,4)\gamma_5 d_e(N_s/2,4) + (u \leftrightarrow d) \} \right] \nonumber \\
\\
\CO_4&=&P^{(G_1)}\left[ \epsilon^{abc}\epsilon^{ade}\epsilon^{bgh}[u_d^T(1,2) C\gamma_5 d_e(1,2)]
	[u_g^T(-1,2) C d_h(-1,2)] C\bar{s}_c^T(0,4) \right] \nonumber \\
\CO_5&=&P^{(G_1)} \left[ \epsilon^{abc}[u_a^T(0,4) C\gamma_5 d_b(0,4)] \right.\nonumber \\ 
      & & \times \left.    \{ u_c(0,4) \left[ \bar{s}_e(N_s/4,4)\gamma_5 d_e(N_s/4,4)
           - \bar{s}_e(-N_s/4,4)\gamma_5 d_e(-N_s/4,4) \right]+ (u \leftrightarrow d) \} \right] \nonumber.
\end{eqnarray}

Here $C$ is the charge conjugation operator and the color indices are shown
explicitly.

The first operator is the one used in our previous work \cite{Csikor:2003ng}
with color index contractions corresponding to an $NK$ state.
$\CO_2$ was introduced in \cite{Sasaki:2003gi}. 
The third operator is a shifted $N-K$ 
scattering operator with spin projection. The relative displacement 
of the nucleon and kaon is half of the spatial lattice size $N_s/2$, so this 
operator is spatially symmetric.
The last two operators are the antisymmetric ones. $\CO_4$ is based on the 
proposal \cite{Jaffe:2003sg}. The two diquarks are shifted to $\pm 1$ from the
origin. Finally, $\CO_5$ is a shifted $N-K$ operator with distance $N_s/4$.
It is first antisymmetrized, then projected to a spin eigenstate.
The projection of the last three operators requires the computation of 
3, 3 and 6 operators, respectively. 
Therefore we have to compute the correlation matrix of
14 elementary operators (except for the elements connecting operators with
opposite spatial symmetry).

\subsection{Simulation parameters and results}

We used the standard Wilson gauge action at $\beta=6.0$ to generate our 
configurations. For the measurements we used the Wilson fermion action with
four different $\kappa_{u,d}$ values for the light 
quarks: 0.1550, 0.1555, 0.1558 and 0.1563. This spans a
pion mass range of 400-630~MeV. For the strange quark we used a constant
$\kappa_{s}=0.1544$, which gives the required kaon mass in the chiral limit.
The lattice size was $24^3\times60$ and for the largest quark mass we also
performed simulations on a $20^3\times 60$ lattice to see the volume 
dependence of the observed states.

\TABLE{
\begin{tabular}{|l||c|r|r|r|} \hline
 size                & operators            & $\kappa_{u,d}$ & confs  
                                                                \\ \hline\hline
   $24^3\times 60$   &  $\CO_1,\CO_2,\CO_3$ & 0.1550   & 242      \\ \hline
   $24^3\times 60$   &  $\CO_1,\CO_2,\CO_3$ & 0.1555   & 205      \\ \hline
   $24^3\times 60$   &  $\CO_1,\CO_2,\CO_3$ & 0.1558   & 205      \\ \hline
   $24^3\times 60$   &  $\CO_1,\CO_2,\CO_3$ & 0.1563   & 205      \\ \hline
\hline
   $20^3\times 60$   &  $\CO_1,\CO_2,\CO_3$ & 0.1550   & 630      \\ \hline
\hline
   $24^3\times 60$   &  $\CO_4,\CO_5$ & 0.1550   & 250      \\ \hline
   $24^3\times 60$   &  $\CO_4,\CO_5$ & 0.1555   & 144      \\ \hline
   $24^3\times 60$   &  $\CO_4,\CO_5$ & 0.1558   & 144      \\ \hline
   $24^3\times 60$   &  $\CO_4,\CO_5$ & 0.1563   & 144      \\ \hline
\hline
   $20^3\times 60$   &   $\CO_4,\CO_5$ & 0.1550   & 234      \\ \hline
   
\end{tabular}
  \caption{The collected statistics for the various simulation points.
      \label{tab:stat}}
}

Table \ref{tab:stat} shows the statistics we collected in the various points.
After performing the spin and parity projections, we used the diagonalization
procedure described in the previous section to separate the possible states in
both parity channels. As mentioned earlier the symmetric operators gave a good
signal only in the negative parity channel while the antisymmetric operators
had reasonable overlap only with positive parity states. This can be 
understood since the parity transformation includes a spatial reflection and
the nucleon-kaon system has a negative inner parity.
Therefore we used only the operators $\CO_1-\CO_3$ to extract negative parity
states and operators $\CO_4-\CO_5$ for positive parity.

We varied both $t$ and $\Delta t$ required 
for the diagonalization over a range of
$2-5$ and included the systematic uncertainties coming from this variation 
in the final errorbars. 
After separating the states we had to extract the lowest masses from the 
individual correlation functions. It turned out that for the excited states
neither a correlated nor an uncorrelated fit with a single exponential (cosh)
was satisfactory since in the asymptotic region where a one exponential
fit could work the data were rather noisy.
We used the following technique instead.

If one plots the effective mass $\log{C(t)/C(t+1)}$ as a function of $t$, it
should show a plateau at asymptotically large $t$ values\footnote{Actually
we used a slightly modified ''effective mass'', namely the
solution of the equation $\cosh(m_{\msub{eff}}\cdot(t-N_t/2))/
\cosh(m_{\msub{eff}}\cdot(t-N_t/2+1))=C(t)/C(t+1)$
to get a flat plateau even for $t$ values close to $N_t/2$.}.
It is easy to show that the effective mass approaches its plateau exponentially:
\be
m_{\msub{eff}}(t)=m+a\cdot\exp(-bt)\;\;\;\;t\rightarrow \infty,
\ee
where $m$ is the lowest mass in the given channel. One can fit the effective
masses with the above formula and use it to extract the lowest masses. In this 
way one also uses the information stored in the points before the plateau
even if the plateau itself is noisy. This technique turned out to be very
stable and we could start to fit the effective masses at $t=2,3$.
Fig.\ \ref{fig:effmass} illustrates the method 
for the first two states in the
negative parity channel for $\kappa=0.1550$.

\FIGURE{
\centerline{\includegraphics[width=12cm]{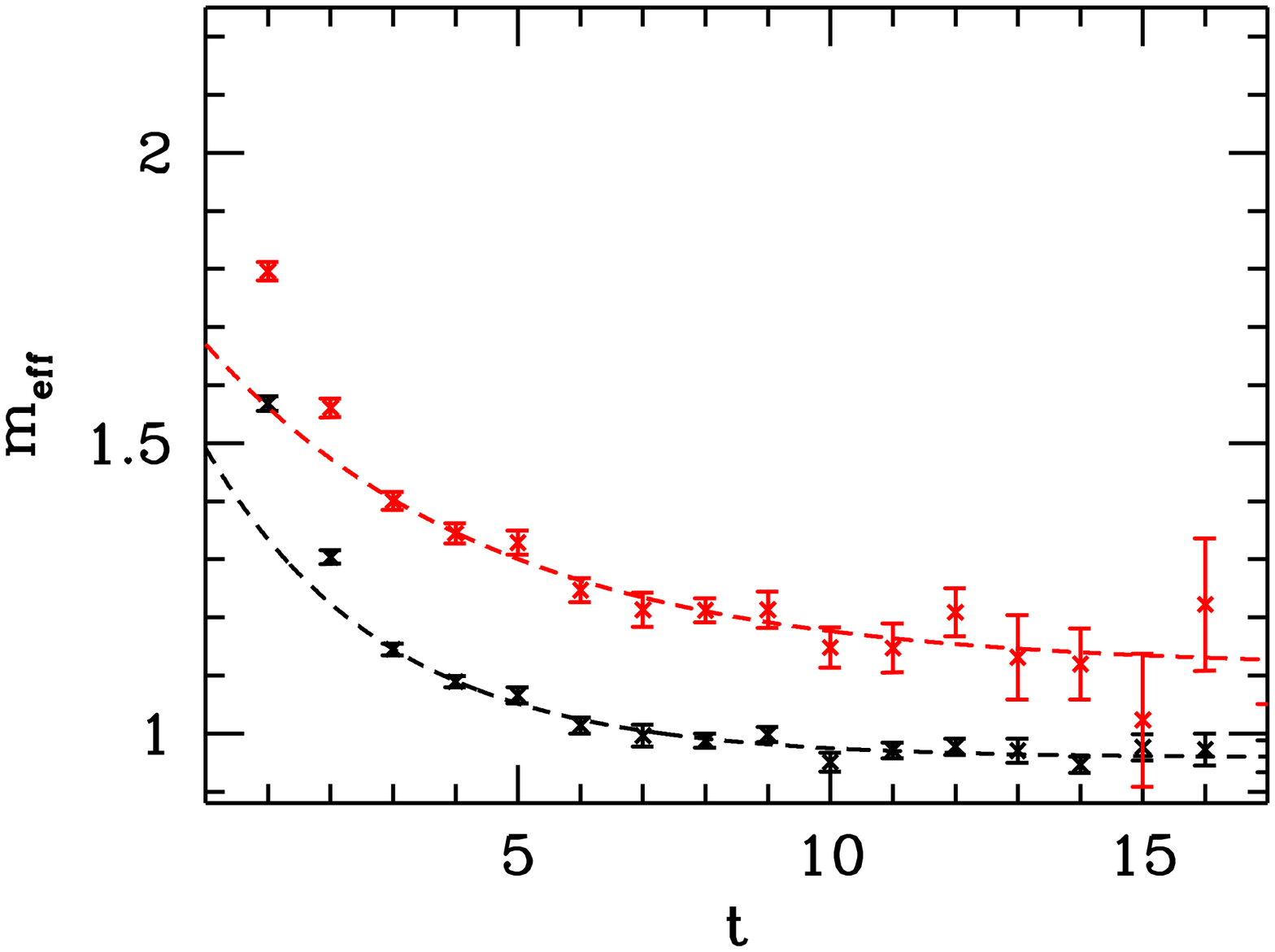}}
\caption{The effective masses for the first two states in the negative
parity channel for $\kappa=0.1550$ and the fitted exponentials.
    \label{fig:effmass}}
}

In both parity channels we extracted the two lowest
masses (which we denote by $m_0$ and $m_1$). It is straightforward to define
the ratio $\alpha_i=m_i/(m_N+m_K)$ which compares the possible scattering
and pentaquark states to the nucleon-kaon threshold. The experimental value
of $\alpha$ for the $\Theta^+$ particle is $\alpha_{\Theta^+}=1.07$.

\TABLE{
\begin{tabular}{|l||c|r|r|r|r|r|r|} \hline
 size & parity & $\kappa_{u,d}$ & $am_{\pi}$ & $am_K$ & $am_N$ & $\alpha_0$ & $\alpha_1$  
                                                                \\ \hline\hline
$24^3\times 60$& -- & 0.1550 & 0.296(1) & 0.317(1) & 0.642(6) & 1.01(1) &1.16(5)\\ \hline
$24^3\times 60$& -- & 0.1555 & 0.259(1) & 0.301(1) & 0.613(5) & 0.99(1) &1.16(5)\\ \hline
$24^3\times 60$& -- & 0.1558 & 0.234(1) & 0.290(1) & 0.592(6) & 0.99(1) &1.14(8)\\ \hline
$24^3\times 60$& -- & 0.1563 & 0.185(1) & 0.272(1) & 0.545(7)& 0.98(2) &1.28(13)\\ \hline
\hline
$20^3\times 60$& -- & 0.1550 & 0.295(1) & 0.316(1) & 0.647(7) & 1.00(1) &1.24(8)\\ \hline
\hline
$24^3\times 60$& + & 0.1550 & 0.295(1) & 0.316(1) & 0.636(6) & 1.16(2) &1.45(16)\\ \hline
$24^3\times 60$& + & 0.1555 & 0.258(2) & 0.299(2) & 0.615(14) & 1.13(3) &1.39(15)\\ \hline
$24^3\times 60$& + & 0.1558 & 0.233(2) & 0.288(2) & 0.595(12) & 1.09(5) &1.39(17)\\ \hline
$24^3\times 60$& + & 0.1563 & 0.184(3) & 0.270(2) & 0.552(10)& 1.14(8) &1.31(32)\\ \hline
\hline
$20^3\times 60$& + & 0.1550 & 0.295(1) & 0.316(1) & 0.647(6) & 1.21(2) &1.48(12)\\ \hline
   
\end{tabular}
  \caption{The measured pion, kaon and nucleon masses and the ratio of the
  first two five-quark states in both parity channels to the $KN$ threshold.
  \label{tab:res}}
}

The summary of our results including
also the pion, kaon and nucleon masses is given in Table \ref{tab:res}.
The zero momentum scattering state is just at the threshold.
The first scattering state with nonzero momentum is expected at
\be
E_1=\sqrt{m_K^2+4\pi^2/(a N_s)^2}+\sqrt{m_N^2+4\pi^2/(a N_s)^2}.
\ee
Its ratio to the threshold is 1.151, 1.166, 1.177, 1.202
for $\kappa=0.1550,0.1555,0.1558$ and $0.1563$, respectively for our larger
volume.
For the smaller volume ($N_s=20$) at $\kappa=0.1550$ this ratio is $1.211$.
We can see that in all cases the measured mass ratios are consistent
with the scattering states. 
The expected and measured volume dependences of the first excited 
state for negative parity and the ground state for positive parity is
shown in Fig. \ref{fig:vol}.

\FIGURE{
\centerline{\includegraphics[width=8cm]{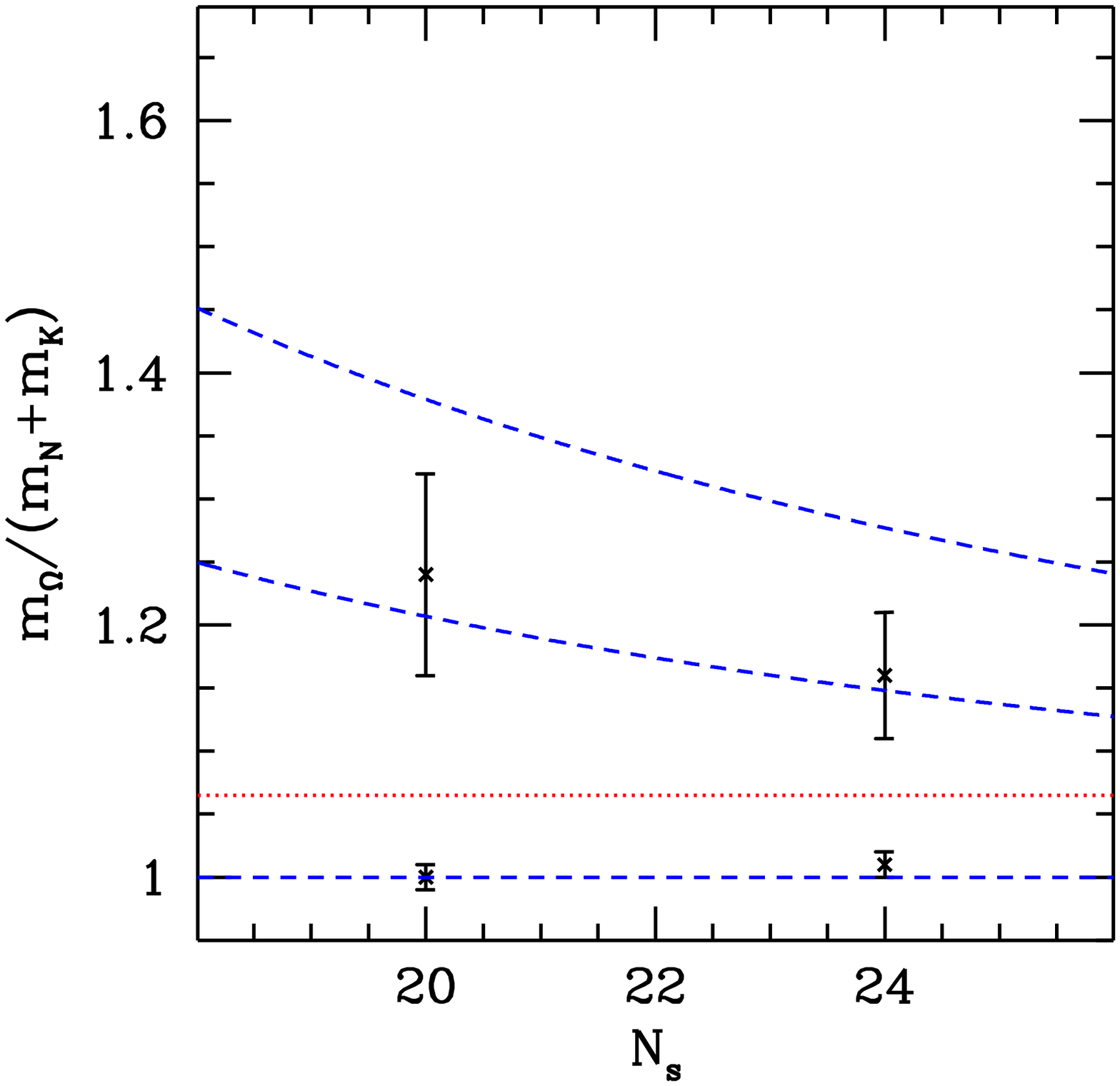}
\includegraphics[width=8cm]{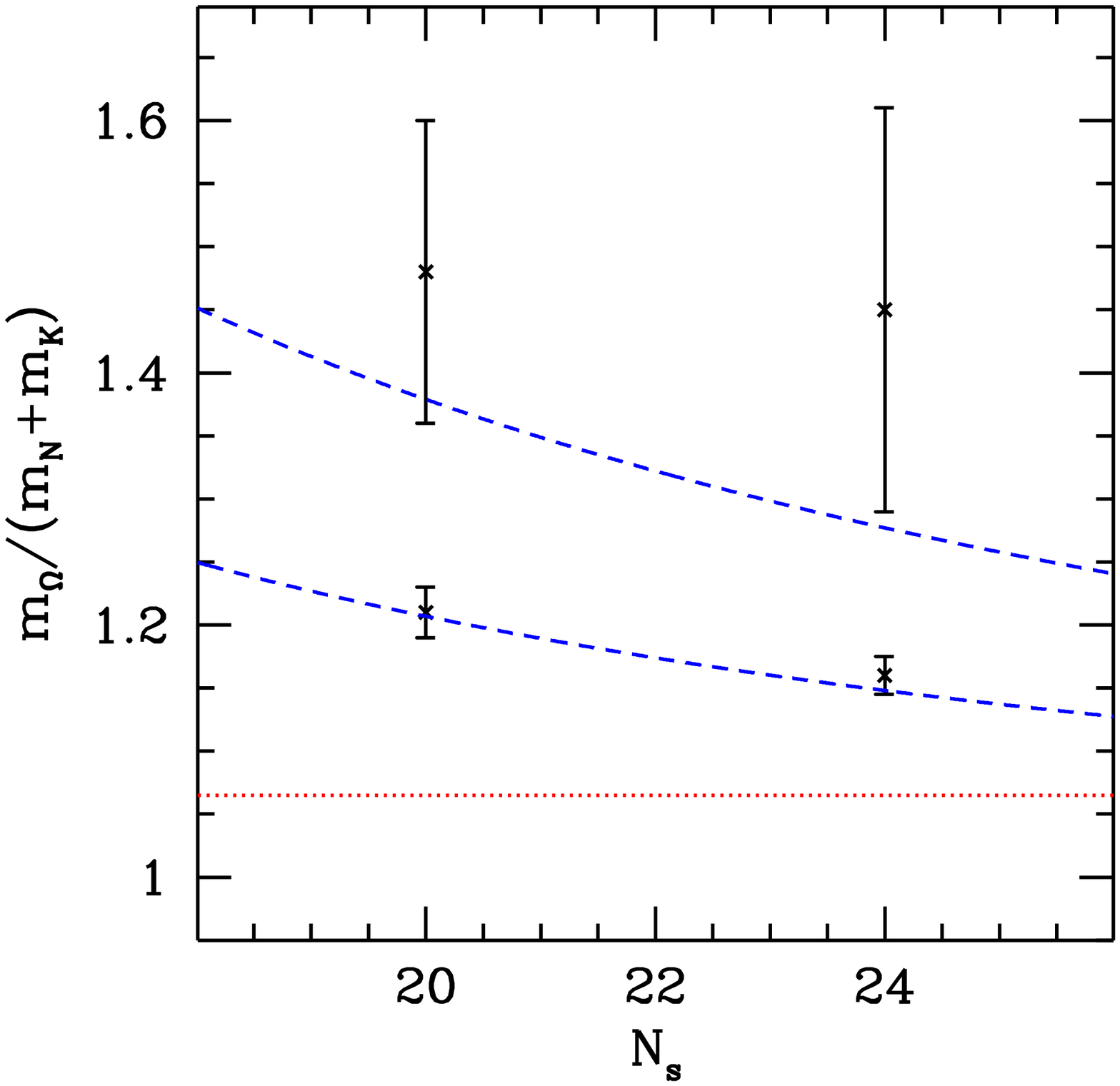}}
\caption{The volume dependence of the two lowest states 
in the two parity channels (left panel: negative parity; 
right panel: positive parity). The dashed lines indicate the
expected scattering states with 0 momentum and the first two 
non-vanishing momenta. The dotted line shows the experimental value
of the pentaquark state.
    \label{fig:vol}}
}

For the highest quark mass, where we had the largest statistics
quark mass we also performed the whole analysis for the isovector channel. The 
extracted masses and their volume dependence turned out to be 
qualitatively similar to those in the isoscalar channel.

\section{Conclusion}

In this paper we studied spin 1/2 isoscalar and isovector, even and
odd parity candidates for the $\Theta^+$ pentaquark using large scale
lattice QCD simulations. The analysis needed approximately  
0.5 Tflopyears of sustained 32 bit operations.

Before we summarize the results of the different channels one technical 
remark is in order. 
The $\Theta^+$ pentaquark is expected to be
a few \% above the NK threshold. Typical lattice sizes
of a few fermis result in a discrete NK scattering spectrum, with order 
10\% energy difference between the lowest lying states. Thus, we are 
faced with two problems. First of all we have to find the possible pentaquark 
signal and the nearby scattering states. Secondly we have to tell the 
difference between them. 
Finding several states could be done by multi-parameter fitting\footnote{
Note that the energies of the states are determined by the exponential
decays of correlation functions; extracting several decay rates, which
differ just by a few \%, from the sum of noisy decays is in practice 
not feasible.} 
or more effectively by spanning a multidimensional wave function
basis and using a cross correlator technique. 
The most straightforward way to tell the difference
between a narrow resonance and a scattering state is to use the fact 
that the former has an energy with quite weak volume dependence, 
whereas the latter has a definite volume dependence, defined by the
momenta allowed in a finite system.

Clearly, any statement on the existence/non-existence or on the
quantum numbers of the $\Theta^+$ pentaquark depends crucially on 
this sort of separation. 
None of the previous lattice investigations on the $\Theta^+$ pentaquark
was able to carry out this analysis. The most important goal of
the present paper was to do it. 

The individual results in 
the odd and even parity channels can be  
summarized as follows (based on the statistically most 
significant, highest quark mass
and assuming that $m_\Theta^+/(m_N+m_K)$ scales with the 
quark mass). 

1. {\it Odd parity.} The two lowest lying states are 
separated. The lower one is identified as the lowest scattering state 
with appropriate volume dependence (in this case the p=0 scattering 
means no volume dependence). This state is  6$\sigma$ below 
the $\Theta^+$ state. The volume dependence of the  second lowest state 
is consistent with that of a scattering state with non-zero relative
momentum. For our larger/smaller volumes this state is 
1.8/1.3$\sigma$ above the $\Theta^+$ state. None of these
two states could be interpreted as the $\Theta^+$ 
pentaquark\footnote{The volume dependence and  
the larger operator basis of this work suggest, 
that the odd parity signal of our previous analysis 
\cite{Csikor:2003ng}, which was quite close to the 
$\Theta^+$ pentaquark mass, was most probably a mixture of 
the two lowest lying scattering states.}.  

2. {\it Even parity.} The two lowest lying states are
identified. The volumes are chosen such that 
even the lowest lying scattering state is above the expected 
$\Theta^+$ pentaquark state. 
The volume dependence of the lowest state suggests that
it is a scattering state. For both volumes this state is
6$\sigma$ above the $\Theta^+$ state.
Since the energy of the second lowest state is even larger, 
none of them could be interpreted as the $\Theta^+$ pentaquark. 

To summarize in both parity channels we identified
all the nearby states both below and above the expected $\Theta^+$
state. Having done that no additional resonance state was found. 
This is an indication that in our wave function basis 
no $\Theta^+$ pentaquark exists (though
it might appear in an even larger, more exotic basis, with smaller
dynamical quark masses or approaching the continuum limit).

\appendix
\section{Parity projection}

In this appendix we summarize how the energy of the lowest state can be
extracted separately in the two parity channels. Although we consider 
spin $1/2$ baryon correlators here, our discussion can be generalized
to other states. 

The basic objects one can compute on the lattice are
Euclidean correlators of the form 
$\langle\,  \CO_\alpha(x) \bar{\CO}_\beta(0) \,\rangle$
corresponding to the amplitude of the process of creating a state
at time zero with the operator $\bar{\CO}_\beta(0)$, evolving it to a
later time $x_0$ and annihilating it with $\CO_\alpha(x)$. 

There are two complications when one wants to extract
the lowest state in a given parity channel. Firstly, simple baryonic
operators usually couple to both parities, therefore one has to project
out parity by hand. Secondly, the box has a finite time extent $T$
with (anti)periodic boundary condition. Therefore,
a single source at time zero is in fact mathematically equivalent to
the sum of an infinite number of identical sources located at 
$t=0,\pm T,\pm2 T...$. Due to the exponential fall-off of correlations, 
only the two sources closest to the sink, i.e.\ at $t=0,T$ give
appreciable contributions to the infinite sum. If we assume, as we shall
always, that $0\le x_0 < T$ then 
\begin{equation}
 \sum_{n=-\infty}^\infty \langle\, \CO_\alpha(\vec{x},x_0) 
      \bar{\CO}_\beta(\vec{0},nT) \,\rangle \approx 
  \langle \, \CO_\alpha(\vec{x},x_0) \bar{\CO}_\beta(\vec{0},0) \,\rangle +
  \varepsilon\msub{bc}
  \langle \, \CO_\alpha(\vec{x},x_0) \bar{\CO}_\beta(\vec{0},T) \,\rangle,
    \label{eq:corr_p-ap}
\end{equation}
where $\varepsilon\msub{bc}$ is $+1$ for periodic and $-1$ for
anti-periodic boundary condition in the time direction.
The first term on the r.h.s.\ represents particles propagating from 
time $0$ to $x_0$ while the second term represents antiparticles propagating
from time $x_0$ to $T$. Thus even after projecting to a given parity
channel the correlator has contributions not only from particles
of that parity, but also from the antiparticles of particles of the 
opposite parity.  Therefore, an additional ``projection'' is needed to get
rid of the latter.

Before starting to describe in detail how the two projections can be 
carried out let us discuss the form of the first term on the r.h.s.\ 
of eq.\ (\ref{eq:corr_p-ap}). We shall assume that $\CO_\alpha$ is a 
spin $1/2$ baryon operator and $\alpha$ is its Dirac index. Due to the 
transformation properties of $\CO_\alpha$, the most
general form the correlator can have is
\begin{equation}
 \langle \, \CO_\alpha(x) \bar{\CO}_\beta(0)\,\rangle =
 \left[ \;  f(x^2) \, x_\mu \gamma^\mu + 
            g(x^2) \, {\bf 1}  \; \right]_{\alpha\beta},
\end{equation}
where $f$ and $g$ are scalar functions of the length of the
four-vector $x$ and ${\bf 1}$ is the $4\times 4$ unit matrix.
After projection to the zero momentum sector this becomes
\begin{equation}
 C_{\alpha\beta}(x_0) = 
 \int\! \mbox{d}^3\!x \; \langle \, \CO_\alpha(x) \bar{\CO}_\beta(0)\,\rangle =
 \left[\, A(x_0)\gamma_0 + B(x_0){\bf 1} \, \right]_{\alpha\beta},
    \label{eq:zeroP}
\end{equation}
where
\begin{equation}
 A(x_0) = \int\! \mbox{d}^3\!x \; f(x^2)x_0, \hspace{1cm}
 B(x_0) = \int\! \mbox{d}^3\!x \; g(x^2).
    \label{eq:AB}
\end{equation}
The important point here is that upon integration over 3-space the
terms in the correlator proportional to the spacelike $\gamma$ 
matrices vanished due to their antisymmetry. We also note that
from eq.\ (\ref{eq:AB}) $A(x_0)$ and $B(x_0)$ are easily seen to
be anti-symmetric and symmetric respectively.

{\em Parity projection.} Parity projection of an arbitrary operator
$\CO$ can be performed by 
\begin{equation}
 {\cal P}_{\pm} \CO = \frac{1}{2}\left(\CO \pm P\CO P^{-1}\right),
\end{equation}
where $P$ is the parity transformation and ${\cal P}_{\pm} \CO$
couples to parity eigenstates of $\pm$ parity. In particular, on
spin $1/2$ fermionic operators parity acts as
\begin{equation}
 P\CO_\alpha(x_0,{\bf x}) P^{-1} = 
 \eta\, (\gamma_0)_{\alpha\nu}\, \CO_\nu(x_0,-{\bf x}) ,
\end{equation}
where $\eta=\pm 1$ is the internal parity of $\CO$ depending on
the parity convention we choose for the elementary fields and on how
$\CO$ is constructed from those. The analogous formula can be easily
obtained for $\bar{\CO}_\alpha$. 

When constructing correlators it is
enough to project to a given parity either at the sink or at the source.
This by itself ensures that only states of the given parity are
propagating in the correlator. Inserting the projection into the correlator of
eq.\ (\ref{eq:zeroP}) at the sink one obtains e.g.\ for the positive 
parity correlator
\begin{equation}
 \frac{1}{2} \left( {\bf 1} + \eta \gamma_0 \right)
 \left[\, A(x_0)\gamma_0 + B(x_0){\bf 1} \, \right] =
  \frac{\eta}{2} \left[ A(x_0) +  \eta B(x_0) \right]{\bf 1}    
 \; + \; \frac{1}{2} \left[      A(x_0) + \eta B(x_0) \right]\gamma_0.
\end{equation}
The negative parity channel can be constructed analogously by replacing 
$\eta$ with $-\eta$ everywhere.

Notice that all the matrix elements of the parity projected correlator
have the same functional dependence, $1/2(A\pm\eta B)$, on $x_0$.
The exponential fit to this function will yield the lowest state in the
given parity channel. In practice the simplest way to obtain the parity 
projected correlator is to compute two suitable elements of the 
$4\times 4$ correlation matrix $C_{\alpha\beta}(x_0)$ that yield $A(x_0)$ 
and $B(x_0)$ respectively. So far we pretended that
the box is infinite in the time direction and neglected the 
second term on the r.h.s.\ of eq.\ (\ref{eq:corr_p-ap}). 

{\em ``Particle'' projection:} Now the full correlator, including the
term that ``comes back'' through the time boundary, the only object 
that we can actually compute in a finite box, has the form
\begin{equation}
 C(x_0) + \varepsilon\msub{bc} C(x_0-T) =
 \left[A(x_0) - \varepsilon\msub{bc} A(T-x_0) \right] \gamma_0 \; + \;
 \left[B(x_0) + \varepsilon\msub{bc} B(T-x_0) \right] {\bf 1},
\end{equation}
where we arranged the arguments of $A$ and $B$ to be non-negative
using their (anti)-symmetry.

Were we to use our prescription above for parity projection, i.e.\
the ``$\gamma_0$ component'' of the correlator $\pm\eta\times$ its 
``${\bf 1}$ component'', we would end up with the parity projected 
correlator
\begin{equation}
      \frac{1}{2} \left[ A(x_0) + \eta B(x_0) \right]
\;+\; \frac{\varepsilon\msub{bc}}{2} \left[ -A(T-x_0) + B(T-x_0) \right],
\end{equation}
which, due to an extra minus sign, does not have the simple
functional form $f(x_0)+f(T-x_0)$ that could be fitted with a cosh.
This extra sign, however, can
be easily canceled if we compute the $A$ and the $B$ components 
(the one proportional to $\gamma_0$ and ${\bf 1}$, respectively in 
eq.\ (\ref{eq:zeroP})) with {\em opposite} boundary conditions. 
In that case the parity projected correlator has the form
\begin{equation}
 \frac{1}{2} \left[   A(x_0) + \eta   B(x_0) \right] \; + \;
 \frac{1}{2} \left[ A(T-x_0) + \eta B(T-x_0) \right].
     \label{eq:cosh}
\end{equation}
If $1/2[A(x_0)+\eta B(x_0)]$ is a sum of exponentials corresponding
to the energies of the states in the given channel, (\ref{eq:cosh})
is a sum of cosh's with the same exponents. This is the functional
form we have to use for fitting when extracting masses.

\section{Projection to a spin eigenstate}
    \label{ap:spin_proj}

In this appendix we outline how a specific spin eigenstate can be
projected out from a given lattice hadron operator. After summarizing
the relevant group theoretical principles we discuss  
how our spin 1/2 pentaquark operators were constructed.

When discussing spin on a hypercubic lattice the first problem
is that due to the absence of full $SO(3)$ rotational symmetry
it is not straightforward to assign spin to a lattice energy 
eigenstate. States on the lattice can be classified into 
irreducible representations of the cubic group $O$ or its
double cover $^2O$, not $SO(3)$ and $SU(2)$ as in the continuum.

With the exception of the lowest four representations, when 
restricted to $^2O$, irreducible representations of $SU(2)$
do not remain irreducible.  The spin 0, 1/2, 1 and 3/2 $SU(2)$
representations are the exceptions, these restricted to $^2O$
are equivalent to the irreducible representations $A_1, G_1,
T_1$ and $H$, respectively. Also any
state belonging to an irreducible representation of $^2O$ has
components belonging to several different spin representations
of $SU(2)$. For instance a state in $G_1$ has components
in spin $1/2, 7/2, 9/2...$ $SU(2)$ representations and 
$H$ has components of spin $3/2, 5/2, 7/2...$. 

This means e.g.\ that if on the lattice we find the lowest energy state 
in the $G_1$ representation of $^2O$, we can identify that with
a spin 1/2 state in the continuum, provided all the higher 
spin states contributing to $G_1$, i.e.\ $s=7/2,9/2...$ can 
be assumed to have much higher energy. In this sense, for practical 
purposes, the lowest few representations of $SU(2)$ and $^2O$ can be 
identified as follows: 
\begin{equation}
 0\leftrightarrow A_1, \hspace{3mm}
1/2 \leftrightarrow G_1, \hspace{3mm}
1 \leftrightarrow T_1, \hspace{3mm} 
3/2 \leftrightarrow H. 
\end{equation}

The task we have at hand is thus to construct states belonging
to specific representations of the cubic group $^2O$. This can
be most easily done using the technique of projection operators
that we summarize here for completeness. 
The simple form of the method of projectors we present here can 
be used only when each 
irreducible representation occurs in the decomposition at most once.
Therefore it is essential to know ahead of time
the irreducible representations occurring in a tensor product and
their multiplicities. This can be most easily found using 
group characters. See e.g.\ \cite{Johnson:1982yq} for explicit
formulae and character tables of $O$ and $^2O$.

Let $G$ be a finite 
group, $D^{(r)}_{ij}(g)$ be the matrix elements of its 
irreducible representation $r$ of dimension $d_r$. Let the transformations
$T(g)$ form an arbitrary (not necessarily irreducible) unitary representation
of $G$. We would like to project a specific irreducible representation
$r$ out of the carrier space of the $T(g)$'s. Let us define the transformations
\begin{equation}
 P^{(r)}_{ij} = \frac{d_s}{|G|} \sum_{g \in G} D^{(r)\star}_{ij}(g) \; T(g), 
     \label{eq:Pij} 
\end{equation}
where $|G|$ is the number of elements $G$ has and $^\star$ denotes
complex conjugation.  

It is straightforward to show that if $|\psi \rangle$ is any 
vector belonging to the carrier space of $T(g)$'s then for 
a fixed $j$ the $d_r$ vectors
\begin{equation}
 |\phi_i\rangle = P^{(r)}_{ij}|\psi\rangle, \hspace{3mm} i=1,...d_r
    \label{eq:basis}
\end{equation}
either transform as basis vectors of the irreducible representation
$r$ or they are all zero. For the proof see any standard text on group 
representations, e.g.\ Ref.\ \cite{Barut-Raczka}. Equations (\ref{eq:Pij})
and (\ref{eq:basis}) can be exploited to project out different 
representations of $^2O$ from a given state on the lattice and its 
rotated copies.

In particular, we would like to construct pentaquark states belonging 
to $G_1$ that corresponds to spin 1/2. Although more complicated cases
can also be considered, here we restrict ourselves to the one where
the spin indices of all the quarks but one have been contracted to be
scalars and the total spin of the pentaquark arises by combining the spin
1/2 ($G_1$) of the remaining quark with the orbital angular momentum
of all the constituents. Therefore we have to project $G_1$ out of 
$G_1 \otimes s$, where $s$ is a representation of the cubic group $O$
(not $^2O$!), corresponding to the orbital part. 

In practice $s$ depends on the spatial arrangement of quark sources
and this can be exploited to make things as simple as possible.
Eq.\ (\ref{eq:Pij}) implies that, in general, projection to a specific spin
involves as many terms as the number of elements of the group $^2O$,
i.e.\ 48. The situation, however, is much better if the projection
formula (\ref{eq:basis}) is applied to a state, with an orbital part
having some degree of symmetry under cubic rotations. 
The simplest case is when the
five quark sources all have complete rotational symmetry, i.e.\ 
the orbital part is trivially $s=A_1$. Then all the rotated copies of the 
quark sources are identical, the sum in eq.\ (\ref{eq:Pij})
can be explicitly computed and the projection reduces to projection
to spin up or spin down. The decomposition here is $A_1\otimes G_1=G_1$. 
All the operators used in lattice pentaquark spectroscopy so far fall
into this category.

To explore the possibility of non-zero orbital angular momentum
we have to consider less symmetric quark sources. Another possibility
is to put the antiquark at the origin with a rotationally symmetric wave
function, displace the two pairs of $(ud)$ quarks along a coordinate
axis (say $z$)  keeping the arrangement cylindrically symmetric with 
respect to the $z$ axis. Inspired by the Jaffe-Wilczek 
diquark-diquark-antiquark picture \cite{Jaffe:2003sg}, 
in anticipation of orbital angular momentum 1, we construct 
this state to be antisymmetric with respect to the interchange of the
two displaced quark pairs. Let us call such a state $|\pm z\rangle$.
It is easy to see that the rotated copies of this state span a
three dimensional space carrying the representation $T_1$ of $O$. 
A possible set of basis states is given by $(ud)$ pairs displaced
along the three coordinate axes; $|\pm x\rangle, 
|\pm y\rangle, |\pm z\rangle$. 
This arrangement corresponds to projecting out the spin 
$1/2$ $(G_1)$ component from the decomposition
\begin{equation}
 T_1 \otimes G_1 = G_1 \oplus H.
    \label{eq:t1g1}
\end{equation}
Let us choose $|\psi\rangle = |\!\uparrow\rangle \otimes |\pm z\rangle$
and compute $P_{11}^{(G_1)}|\psi\rangle$. The transformations $T(g)$
appearing in eq.\ (\ref{eq:Pij}) are direct products of $G_1$ transformations
acting on the quark spin and transformations acting on the orbital part.
Each term in the sum and as a consequence the whole sum itself can 
be decomposed into three terms proportional to $|\pm x\rangle, 
|\pm y\rangle$ and
$|\pm z\rangle$. The $G_1$ matrices can be easily obtained by restricting
the defining representation of $SU(2)$ to $^2O$ and with the factors 
$D^{(G_1)\star}_{ij}(g)$ they can be summed independently for the three
terms resulting in
\begin{eqnarray}
P_{11}^{(G_1)}\; \left[ |\!\uparrow\rangle \otimes |\pm z\rangle \right] & = &
{ \left( \begin{array}{cc} 0 & 0 \\
                         1 & 0 
       \end{array}      \right) } |\! \uparrow\rangle\otimes |\pm x\rangle + 
{ \left( \begin{array}{cc} 0 & 0 \\
                         i & 0 
       \end{array}      \right)} |\! \uparrow\rangle\otimes |\pm y\rangle + 
{ \left( \begin{array}{cc} 1 & 0 \\
                         0 & 0 
       \end{array}      \right)} |\! \uparrow\rangle\otimes |\pm z\rangle 
  \nonumber \\
& = &  |\!\downarrow\rangle \otimes |\pm x\rangle + 
      i|\!\downarrow\rangle \otimes |\pm y\rangle + 
       |\!\uparrow\rangle \otimes |\pm z\rangle.
\end{eqnarray} 
In a similar fashion we obtain the other (spin down) basis element of
the $G_1$ projection;
\begin{equation}
  P_{21}^{(G_1)}\; \left[ |\!\uparrow\rangle \otimes |\pm z\rangle \right] 
  \; = \;
        |\!\uparrow\rangle \otimes |\pm x\rangle 
      \; - \; i|\!\uparrow\rangle \otimes |\pm y\rangle  
      \; - \; |\!\downarrow\rangle \otimes |\pm z\rangle.
\end{equation}
Note that, up to some numerical factors coming from the normalization
of spherical harmonics, these expressions are identical to the spin $1/2$
part of the $SU(2)$ Clebsch-Gordan decomposition 
$1\otimes 1/2= 1/2 \oplus 3/2$. We could also construct 
spin 1/2 from similar, but symmetric orbital states
for the quarks displaced to $x,y,z=-d$.  This would correspond to 
$A_1\otimes G_1=G_1$  or $0\otimes 1/2=1/2$ for $SU(2)$. 

Building these states requires seven quark sources; an antiquark at the
origin and six quark sources, two along each coordinate axis (we use the same
mass and source for the $u$ and $d$ quarks). Eq.\ (\ref{eq:t1g1}) shows 
that keeping the same spatial arrangement the representation $H$ 
corresponding to spin $3/2$ could also be projected out. However,
we have not explored this possibility here. For that
we would have had to replace the matrix elements $D^{(G_1)\star}_{ij}(g)$
in eq.\ (\ref{eq:Pij}) with those of $H$. 

Besides the diquark-diquark-antiquark wave function we also wanted to 
study triquark quark-antiquark states. The simplest non-trivial way
to do that is to displace the quark-antiquark pair along a coordinate
axis, say $+z$. Let us call the orbital part of this state $|+z\rangle$.
Its rotated copies span the six dimensional space with a possible
basis formed by $|+x\rangle, |-x\rangle, |+y\rangle,  |-y\rangle, |+z\rangle,
|-z\rangle$. This space, however, can be split into an antisymmetric part
spanned by combinations of the form $|+x\rangle-|-x\rangle,...$ and
a symmetric one spanned by $|+x\rangle+|-x\rangle,$ etc. The 
representation of $O$ on the antisymmetric part is $T_1$ in exactly
the same way as in the diquark-diquark case, resulting again in the 
spin projected state
\begin{eqnarray}
   P_{11}^{(G_1)}\; \left[ |\!\uparrow\rangle \otimes 
                 (|+z\rangle -|-z\rangle)\right] = \hspace{5cm}
   \nonumber  \\
      |\!\downarrow\rangle \otimes (|+x\rangle -|-x\rangle)
      \; + \; i|\!\downarrow\rangle \otimes (|+y\rangle -|-y\rangle)
      \; + \; |\!\uparrow\rangle \otimes (|+z\rangle -|-z\rangle). 
\end{eqnarray}

The three dimensional symmetric part of the orbital space is 
reducible to $A_1\oplus E$. 
Thus we can also produce spin 1/2 trivially from the symmetric part 
by $A_1\otimes G_1=G_1$,
\begin{eqnarray}
   P_{11}^{(G_1)}\; \left[ |\!\uparrow\rangle \otimes 
                 (|+z\rangle +|-z\rangle)\right] = \hspace{5cm} \nonumber \\
 |\!\uparrow\rangle \otimes 
       \left[ |+x\rangle +|-x\rangle +
        |+y\rangle +|-y\rangle +
        |+z\rangle +|-z\rangle \right].
\end{eqnarray}

\acknowledgments

This research was partially supported by
OTKA Hungarian Science Grants No.\ T34980, T37615, M37071, T032501.
This research is part of the EU Integrated Infrastructure Initiative 
Hadronphysics project under contract number RII3-CT-20040506078.
The computations were carried out at E\"otv\"os University 
on the 330 processor PC cluster of the Institute for Theoretical Physics 
and the 1024 processor PC cluster of Wuppertal University, using
a modified version of the publicly available MILC code (see 
www.physics.indiana.edu/ \~{ }sg/milc.html). T.G.K.\ also acknowledges 
support through a Bolyai Fellowship.

\end{document}